\documentclass[twocolumn,showpacs,prl,floatfix]{revtex4}
\usepackage{graphicx}
\usepackage{amsmath}
\usepackage{dcolumn}
\usepackage{bm}

\newcommand{\Fig}[1]{Figure~\ref{fig:#1}}

\newcommand{\Eq}[1]{Eq.\ (\ref{eq:#1})}
\newcommand{\Eqs}[2]{Eqs~(\ref{eq:#1})~and~(\ref{eq:#2})}

\usepackage{latexsym}

\begin{document}

\title{Potential Vorticity Formulation of Compressible Magnetohydrodynamics}
\author{Wayne Arter}
\affiliation{EURATOM/CCFE Fusion Association, Culham Science Centre, Abingdon, UK. OX14 3DB}
\date{\today}

\begin{abstract}
Compressible ideal magnetohydrodynamics (MHD) is formulated in terms of the 
time evolution of potential vorticity and magnetic flux per unit mass
using a compact Lie bracket notation.
It is demonstrated that this simplifies analytic solution
in at least one very important situation relevant to magnetic fusion
experiments. Potentially important implications
for analytic and numerical modelling of both laboratory
and astrophysical plasmas are also discussed.
\end{abstract}

\pacs{52.30.Cv, 52.55.Fa, 96.60.Q-}

\maketitle

\section{Introduction}\label{sec:intro}
Ideal MHD is a model for magnetised plasma where the collisionality is low,
so that dissipative effects can be neglected, yet where the charged
particles still interact sufficiently strongly via the electromagnetic
field they can be treated as a single fluid.
The ideal MHD model is applied to
a wide range of laboratory and astrophysical situations, where there are
long periods of relative quiescence in which Maxwellian particle
distributions can be approached, interrupted by often violent transients.
Ideal MHD instabilities are thought to be implicated in the triggering
of the sawtooth crash phenomenon in tokamak magnetic fusion experiments
and flaring in the solar and stellar context, see textbooks such
as~\cite{goedbloedpoedts}. The former is important
as it limits the performance of devices ultimately intended to generate
nuclear power, and the latter is implicated in the generation of
solar magnetic storms which can disrupt terrestrial power grids,
navigation and communication systems. Both these topics are presently
the subject of intensive investigation, magnetic fusion
as the multi-billion dollar ITER tokamak enters the construction phase,
whereas multiple satellite missions are collecting
data on solar and stellar magnetic fields.

It is often mathematically convenient
when employing ideal MHD, to assume
that the plasma fluid is \emph{incompressible}, but the reality in the
above-mentioned situations is that the plasma density varies by
one or more orders of magnitude over the region of interest.
This work presents what is believed to be a novel, mathematically
convenient formulation of \emph{compressible} MHD.

The equations of ideal MHD as usually formulated are well-known
and are to be found in many textbooks, see eg.\ \cite[\S\,4.3]{goedbloedpoedts}.
As explained there, the problem admits a variational formulation
which is of great utility for practical stability analysis,
and a functional Hamiltonian formulation in terms of Lie
derivatives~\cite{Mo80Nonc}, of great theoretical importance for
understanding stability and evolution. More direct approaches
to ideal MHD stability are also now used~\cite[\S\,6]{goedbloedpoedts},
and the results presently to be described are more relevant to
the latter school.

The potential vorticity is the ordinary vorticity~$\boldsymbol{\omega}$
of the plasma
(the curl of the mean flow~${\bf U}$ of ions and electrons),
divided by the mass density~$\rho$, ie.\ $\tilde{\boldsymbol{\omega}}=\boldsymbol{\omega}/\rho$.
The possibility of combining the equation for the time evolution of
vorticity with that for density evolution to give a simple
equation for the rate of change of potential vorticity, was first
realised for a classical fluid by Helmholtz as described by~\cite[\S\,146]{lamb}
in the mid-19th Century. In the mid-20th Century, Wal\'en, according 
to~\cite[\S\,4-2]{boydsanderson} was the first to realise that a mathematically
identical relation governed the evolution of the magnetic
flux per unit mass~$\tilde{\bf B}={\bf B}/\rho$ where ${\bf B}$ is the magnetic field.
For \emph{incompressible} plasma, Arnold \& Khesin~\cite[\S\,I.10.C]{arnoldkhesin}
combined these results in late-20th Century
to give an elegant formulation of
ideal MHD in terms of Lie brackets of vector fields. The Lie bracket
is here the generalisation to arbitrary vector fields of the 
`flux-freezing' operator, ie.\ the operator which determines the advection of divergence-free
(solenoidal) fields~${\bf B}$ and $\boldsymbol{\omega}$~\cite[\S\,3.8]{schindler}. The
novelty of the present work is to extend this formalism
to \emph{compressible} MHD and explore the implications. In particular,
the peculiar, coordinate invariant nature of the Lie bracket makes it
easy to generalise solutions to arbitrary geometry in some cases,
both analytically and numerically.

The next section contains a detailed mathematical derivation of
the key formula. A discussion of the implications for analytic and numerical
solution follows, and finally some
important possible applications are summarised.

\section{Mathematics}\label{sec:maths}
In terms of the operators of Classical Vector Mechanics, the Lie
derivative of a vector can be defined as:
\begin{equation}\label{eq:liebr2}
{\bf\mathcal{L}}_{\bf u}({\bf v})=\nabla\times({\bf u}\times{\bf v})-{\bf u}\;\nabla\cdot{\bf v}+{\bf v}\;\nabla\cdot{\bf u}
\end{equation}
which will help explain the equivalence with the vector advection operator,
the first term on the right. Indeed, Wal\'en's result for magnetic
induction in a perfectly conducting medium is
\begin{equation}\label{eq:induc}
\frac{\partial \tilde{\bf B}}{\partial t}={\bf\mathcal{L}}_{\bf U}(\tilde{\bf B})
\end{equation}
Introducing component notation for vectors in general non-orthogonal
coordinate
systems, as described in many textbooks e.g.\ \cite{dhaeseleer}, it turns out that
the Jacobians thereby introduced (of the co-ordinate transformation from Cartesians),
cancel among the terms in \Eq{liebr2}, so that
\begin{equation}\label{eq:lieder}
{\bf\mathcal{L}}_{\bf u}({\bf v})^i= v^k\frac{\partial u^i}{\partial x^k}-u^k\frac{\partial v^i}{\partial x^k}
\end{equation}
where $u^k,v^k$ are the contravariant components of the 3-vectors~${\bf u},{\bf v}$
respectively, and the summation convention is implied. It follows that
\begin{equation}\label{eq:liebr}
{\bf\mathcal{L}}_{\bf u}({\bf v})=-{\bf\mathcal{L}}_{\bf v}({\bf u})=-[{\bf u},{\bf v}]
\end{equation}
where $[\cdot,\cdot]$ denotes the Lie bracket of Schutz~\cite{schutz}.

It will be now be proved that the equation for the evolution of potential
vorticity in compressible ideal MHD may be written
\begin{equation}\label{eq:fullpv}
\frac{\partial \tilde{\boldsymbol{\omega}}}{\partial t}={\bf\mathcal{L}}_{\bf U}(\tilde{\boldsymbol{\omega}})-{\bf\mathcal{L}}_{\tilde{\bf B}}(\tilde{\bf J})
\end{equation}
where the potential current $\tilde{\bf J}=\nabla\times{\bf B}/\rho$.
The customary vorticity equation in ideal MHD is
\begin{equation}\label{eq:vort}
\frac{\partial\boldsymbol{\omega}}{\partial t}=
\nabla\times\left({\bf U}\times\boldsymbol{\omega}\right)+\frac{\nabla\rho\times\nabla p}{\rho^{2}}+
\nabla\times{\left(\frac{{\bf J}\times{\bf B}}{\rho}\right)}
\end{equation}
where vorticity $\boldsymbol{\omega}=\rho\tilde{\boldsymbol{\omega}}=\nabla\times{\bf U}$, and
current~${\bf J}=\rho\tilde{\bf J}=\nabla\times{\bf B}=\nabla\times(\rho\tilde{\bf B})$.
When proceeding further, it is convenient and often physically justifiable,
by a barotropic or isentropic assumption, to neglect the term in the pressure~$p$,
and if not, the resulting additional term is easily representable
in general geometry.

It follows that to establish the equivalence of \Eqs{fullpv}{vort},
it is necessary to show
that $\boldsymbol{\Delta}={\bf 0}$, where
\begin{equation}\label{eq:thm}
\boldsymbol{\Delta}=\frac{1}{\rho}\nabla\times{\left(\frac{{\bf B}\times{\bf J}}{\rho}\right)}-{\bf\mathcal{L}}_{\tilde{\bf B}}(\tilde{\bf J})
\end{equation}

Now, \Eq{thm} is a vector equation, so validity in any coordinate frame implies
validity in all, hence it is sufficient to establish the result in Cartesian
coordinates, where
\begin{equation}\label{eq:thm0}
\boldsymbol{\Delta}=\frac{1}{\rho}\nabla\times{\left(\rho{\tilde{\bf B}\times\tilde{\bf J}}\right)}
+\tilde{\bf B}\cdot\nabla\tilde{\bf J}-\tilde{\bf J}\cdot\nabla\tilde{\bf B}
\end{equation}
The curl term may be expanded using the identity
\begin{equation}\label{eq:thm2}
\frac{1}{\rho}\nabla\times(\rho {\bf v})={\bf R} \times {\bf v}+ \nabla\times{\bf v}
\end{equation}
where ${\bf R}= {\nabla \rho}/{\rho}$. Setting ${\bf v}=\tilde{\bf B}\times\tilde{\bf J}$,
and expanding the resulting curl-cross operation, there is cancellation of the
two terms from the Lie derivative, leaving
\begin{equation}\label{eq:thm3}
\boldsymbol{\Delta}=
\tilde{\bf B}\nabla\cdot\tilde{\bf J}-\tilde{\bf J}\nabla\cdot\tilde{\bf B}
+{\bf R}\times(\tilde{\bf B}\times\tilde{\bf J})
\end{equation}
Since $\nabla\cdot{\bf J}= 0$, it follows that 
\begin{equation}\label{eq:divm}
\nabla\cdot\tilde{\bf J}=-{\bf R}\cdot \tilde{\bf J}
\end{equation}
and likewise since $\nabla\cdot{\bf B}= 0$,
\begin{equation}\label{eq:divh}
\nabla\cdot\tilde{\bf B}=-{\bf R}\cdot \tilde{\bf B}
\end{equation}
Substituting \Eq{divm} and \Eq{divh} in \Eq{thm3}, and expanding the last term
as dot products, shows that, as required $\boldsymbol{\Delta}= {\bf 0}$.

The set of evolution equations is completed by mass conservation
\begin{equation}\label{eq:mconsv}
\frac{\partial \rho }{\partial t}=-\nabla\cdot (\rho {\bf U} )
\end{equation}
This does not involve a vector Lie derivative, but,
using the standard expression for the divergence operator in general
curvilinear coordinates, it may be written
\begin{equation}\label{eq:mconscpt}
\frac{\partial \rho }{\partial t}=-\frac{1}{\sqrt{g}}\frac{\partial (\rho \sqrt{g} U^k )}{\partial x^k}
\end{equation}
where $\sqrt{g}$ is the Jacobian and the $g_{ik}$ is the metric tensor,
which upon introducing $\tilde{\rho}=\rho \sqrt{g}$ may be written
\begin{equation}\label{eq:mcons}
\frac{\partial \tilde{\rho} }{\partial t}=-\frac{\partial (\tilde{\rho} U^k )}{\partial x^k}
\end{equation}
provided that~$\sqrt{g}$ does not change with time. Like the neglect of
the pressure term above, this latter inessential assumption is often physically reasonable.

Unfortunately, the ideal MHD equations are here completed by the two
definitions of potential
vorticity and potential current, which \emph{do} explicitly contain metric
information, viz.
\begin{equation}\label{eq:defpv}
\tilde{\rho} \tilde{\omega}^i = e^{ikl} \frac{\partial (g_{ln} U^n )}{\partial x^k}
\end{equation}
and 
\begin{equation}\label{eq:defpj}
\tilde{\rho} \tilde{J}^i = e^{ikl} \frac{\partial}{\partial x^k} \left(
\frac{g_{ln}}{\sqrt{g}}  \tilde{\rho}  \tilde{B}^n \right)
\end{equation}
In the above, $e^{ikl}=e_{ikl}$ is the alternating symbol, taking values $1$, $-1$ or~$0$,
depending whether $(ikl)$ is an even, odd or non-permutation of $(123)$.
Finally, note that \Eq{induc} and \Eq{mcons} together ensure that $\nabla\cdot {\bf B}=0$,
only if initially
\begin{equation}\label{eq:divh0}
\frac{\partial (\tilde{\rho} \tilde{B}^k )}{\partial x^k}=0
\end{equation}

\section{Solving the New System}\label{sec:solve}
The new model system for ideal barotropic compressible MHD evolution
consists of \Eq{fullpv}, \Eq{induc}, \Eq{mcons}, \Eq{defpv} and \Eq{defpj}.
The simplification of the first three has been gained at the expense
of complicating the last two `static' relations. Nonetheless, evolution
equations are harder to treat numerically, because any errors in the
discretisation tend to combine over time. Moreover, it will be
evident that problems solved in Cartesian geometry will test all aspects
of the coding of the evolutionary equations. Thus, there is considerable computational
advantage to be gained. There is obviously the concern that the magnetic
field computed may not be accurately solenoidal, but this is an issue
for many other discretisations also. The main difficulty is in the
inversion of \Eq{defpv} to give the velocity field~${\bf U}$
corresponding to
a freshly evolved potential vorticity (since $\tilde{\bf B}$ itself is evolved, 
\Eq{defpj} does not need to be inverted). However, this inversion, together with
the computation of the irrotational part of~${\bf U}$, is a classical
hydrodynamical problem, and a variety of strategies may be found in
the literature. On present machine architectures, introducing
the vector potential for velocity then solving 
the coupled system \Eq{mcons} and \Eq{defpv} by a
pseudo-timestepping algorithm is probably to be preferred. Similar
numerical solution strategies were successfully employed in electromagnetics
by the current author and collaborators~\cite{Wa95d,Wa02}. Vorticity
formulations are common in plasma modelling as they are helpful
in several physically relevant limits, and in particular,
a vorticity formulation has been used successfully in 
nonlinear, compressible MHD~\cite{Ch90Compwarv}.

Turning to analytic results, first consider MHD equilibrium solutions 
with no time dependence and ${\bf U}={\bf 0}$, implying
${\bf\mathcal{L}}_{\tilde{\bf B}}(\tilde{\bf J})=0$.
In the case of force-free fields, meaning ${\bf J}\propto{\bf B}$,
substituting $\tilde{\bf J}=\lambda\tilde{\bf B}$
in the Lie derivative in component form show this is a solution
provided ${\bf B}\cdot{\nabla \lambda}= 0$, i.e. exactly the same constraint
on~$\lambda$ that follows from the solenoidal constraint on~${\bf B}$ 
and~${\bf J}$ when
seeking the solution ${\bf J}=\lambda{\bf B}$.
Hydromagnetic force-free solutions, with the additional constraint that
${\bf U}=\lambda_2{\bf B}$, now
cease to exist however, because ${\bf U}$ is not
solenoidal unless the flow is incompressible.

\begin{figure}
\centerline{\rotatebox{270}{\includegraphics[width=5.5cm]{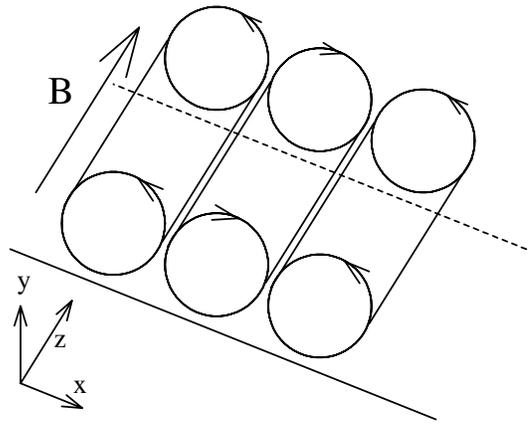}}}
\caption{An, in effect unmagnetised, compressible flow is shown. The
motion consists of 
rolls swirling about a ${\bf B}$-field aligned with the $z$-axis,
with arrow-heads indicating the sense of motion of each eddy.
\label{fig:roll2}}
\end{figure}

Moving now to time dependent solutions, interest attaches to the `flux compression'
solution~\cite[\S\,4.6]{artsimovich}, which is postulated on purely kinematic grounds
(i.e.\ from \Eq{induc}) and which may be written
\begin{equation}\label{eq:Kuls}
{\bf B}=c\left(0,0,\rho(x,y,t)\right)
\end{equation}
for a compressible flow~${\bf U}$ with density~$\rho$ provided 
that ${\bf U}=\left(U_x(x,y,t),U_y(x,y,t),0\right)$. Here, $c$ is an
arbitrary constant and $(x,y,t)$ are the usual Cartesian coordinates.
This solution is of practical importance for fusion experiments,
where external magnets are used to generate a time dependent flux
designed so as to compress plasma `frozen' to it.
It is easy to establish that if ${\bf B}=c\rho \hat{\bf z}$ then
${\bf J}\times{\bf B}/c^2=(\nabla\rho \times \hat{\bf z}) \times\rho \hat{\bf z}=
-\nabla(\frac{1}{2}\rho^2)$, and so there are compressible MHD solutions
of the form \Eq{Kuls}, for 2-D solutions of compressible hydrodynamics
compatible with an additional pressure gradient of this form.
One possibility is illustrated in simple geometry in \Fig{roll2}.

The simple form of the new evolution equations enables a generalisation of
the flux-compression solution to general curvilinear coordinates. It
is important to emphasise that the following is not simply re-expressing
${\bf B}=\rho \hat{\bf z}$ in different coordinate systems, nor is
there a loss of generality in choosing units for density
such that $c=1$.
The obvious generalisation is to take $\tilde{B}^3=1$ ($\tilde{B}^1=\tilde{B}^2=0$), implying
a 2-D density to ensure a solenoidal~${\bf B}$,
since \Eq{divh0} requires
$\partial \tilde{\rho}/\partial x^3=0$.
The next step is to ensure that ${\bf\mathcal{L}}_{\tilde{\bf B}}(\tilde{\bf J})=0$, which as may be seen
using the coordinate form~\Eq{lieder}, simply requires
$\partial \tilde{J}^j/\partial x^3=0$. Similarly
${\bf\mathcal{L}}_{\bf U}(\tilde{\bf B})=0$ may be satisfied by a flow with
$\partial U^j/\partial x^3=0$ (note that
$U^3 \neq 0$ is therefore allowed). From the `static'
relations, it will be seen that a solution 
with $\tilde{J}^j$ independent of~$x^3$ is possible
provided~$\partial g_{ik}/\partial x^3=0$. Put in the language of
differential geometry~\cite[\S\,3.11]{schutz}, if $\tilde{\bf B}$ is a Killing vector,
there is a flux-compression solution.

Further to explore the implications of this, introduce
generalised toroidal coordinates~$(\varrho, s, w)$ (cf.\ $(r,\theta,\phi)$
as commonly employed in plasma physics~\cite{dhaeseleer}) so that 
\begin{equation}\label{eq:torc}
{\bf x}=(x,y,z)=\left(R_c\cos w, R_c\sin w, \psi_c\sin s\right),
\end{equation}
where 
\begin{equation}\label{eq:rtorc}
R_c=R_0+\psi_c(\varrho,s,w) \cos s
\end{equation}
It will be seen that $\psi_c(\varrho,s,w)=const.$ as $\varrho$~varies
form  a set of nested
toroidal surfaces with major axis~$R_0$. Introduce helical
coordinates~$(u,v)$ on each surface, so that $s=u-v/q(\psi)$,
$w=v+u/q(\psi)$, and write $\psi(\varrho,u,v)=\psi_c(\varrho,s,w)$.
Suppose that $\psi$ is rotationally symmetric about the $z$-axis
and satisfies the Grad-Shafranov equation,
ie.\ $\psi$~is a flux function for an equilibrium magnetic field, then
the curves of \Eq{torc} as $v$ varies at constant $u$~and~$\psi$ are equivalent to lines
of the equilibrium field with helical twist~$q(\psi)$. (Note that $u$ and $v$
need only be suitably periodic functions of the regular toroidal
angles~$\theta$ and~$\phi$. To define an equilibrium fully requires
defining these functions, but this is inessential for what follows.)
The metric tensor in a coordinate system $(x^1,x^2,x^3)$ is given by
\begin{equation}\label{eq:gikd}
g_{ik}=\frac{\partial{\bf x}}{\partial x^i}\cdot\frac{\partial{\bf x}}{\partial x^k}
\end{equation}
Taking $(x^1,x^2,x^3)=(\varrho, u,v)$ and using suffix~$,i$ to denote
differentiation with respect to~$x^i$, the components of~$g_{ik}$
are straightforwardly calculated as
\begin{equation}\label{eq:gik}
g_{ik}=\psi_{,i}\psi_{,k} + \psi^2 s_{,i}s_{,k}
+ (R_0+\psi \cos s)^2 w_{,i}w_{,k}
\end{equation}
and when $q$ is constant, $s_{,i}=(0,1,-1/q)$, $w_{,i}=(0,1/q,1)$.

This $x^k$~coordinate system has been chosen so that the equilibrium
field expected
in the tokamak confinement device may be expressed as $\tilde{B}^3=1$ ($\tilde{B}^1=\tilde{B}^2=0$),
but it will be seen that in general, the metric tensor does depend
on~$x^3=v$ through $s=u-v/q$.
By inspection, however, in the limit when $q$ is large, $g_{ik}$
depends only on $x^1$~and~$x^2$. Hence a purely toroidal field,
ie.\ one tangent to circles about the major axis~$x=y=0$ of the torus,
allows for flux freezing solutions. Further, when $\psi/R_0$ is small,
the $s$-dependence of $g_{ik}$ is weak, so a helical field in a
torus with relatively large major radius is also in this category.
The preceding limits illustrate two of the Killing vector solution
symmetries~\cite[\S\,5.2.4]{schindler} (the third is simply invariance
in a Cartesian coordinate).

Other possibilities for new analytic
solutions outside of $\tilde{B}^3=1$ are opened up
when it is realised that the helical field
considered above is just one example of the use of
Clebsch variables~\cite[\S\,5]{dhaeseleer} to represent a
solenoidal vector field as
a single contravariant component. Alternatively, the vector potential
may be introduced, leading to an interesting calculus involving~${\bf R}$,
consistent with the fact that exponentially varying density profiles 
(implying constant~${\bf R}$) are often studied analytically.

\section{Applications}\label{sec:appl}
For magnetic fusion physics,
the above, new analytic
flux-compression solutions represent a possible nonlinear
development of interchange modes~\cite[\S\,12.1.2]{GKP}.
They would seem to represent an efficient and rapid means
whereby mass (and hence heat) might escape from a discharge,
hence might be implicated in situations where there is rapid
transient cooling, such as the sawtooth crash in
the centre of the tokamak discharge ($\psi$~small), and
ELMs (Edge Localised Modes) in divertor discharges (large~$q$~limit).
The preceding section has also speculated that the new formalism
could be used efficiently to simulate ideal MHD evolution
of discharges in generalised coordinates, say defined by
an arbitrary MHD equilibrium.

In astrophysics, observed magnetic fields usually exhibit a
significant degree of disorder, so it is unclear how important
the new flux-compression solutions might be, as they rely on at least
a degree of coordinate invariance. It is speculated that,
in stars with a strong internal toroidal field (such as the Sun is
believed to possess), the rotationally symmetric 
solution might help model the convection pattern, accounting for the
largely latitudinal variation of the solar differential rotation.
Regardless, it should be helpful that, in the new equations,
the field geometry appears only in
the state equations. It will for example, be simpler
to generate more realistic solutions from symmetric ones by
varying~$g_{ik}$ starting with the unit tensor.
This could be useful, say,
for modelling sunspot penumbrae
both analytically and computationally,
since there the magnetic field is predominantly
directed radially outwards in the horizontal direction.

\section*{Acknowledgement}\label{sec:ackn}
I thank Anthony~J.~Webster and John~M.~Stewart for their various helpful inputs.
This work was funded by the RCUK Energy Programme under grant EP/I501045 and
the European Communities under the contract of Association between EURATOM and CCFE.
The views and opinions expressed herein do not necessarily reflect those of
the European Commission.


\end{document}